\RequirePackage{tikz}
\documentclass[11pt,letterpaper]{article}
\usepackage{graphicx}
\usepackage{pgfplots}
\usepackage{pgfplotstable}
\usepackage{booktabs}
\usepackage{authblk}
\usepackage{amssymb}
\usepgfplotslibrary{statistics, groupplots}
\pgfplotsset{compat=1.17}

\usepackage{amsmath}

\usepackage[eulergreek]{sansmath}
\pgfplotsset{
  tick label style = {font=\sansmath\sffamily},
  every axis label = {font=\sansmath\sffamily},
  legend style = {font=\sansmath\sffamily},
  label style = {font=\sansmath\sffamily},
  title style = {font=\sansmath\sffamily}
}
\tikzset{every picture/.style={/utils/exec={\sffamily}}}

\pgfplotsset{every axis/.append style={
                    xlabel={$x$},          
                    ylabel={$y$},          
                    label style={font=\sffamily},
                    tick label style={font=\sffamily\footnotesize},
                    xticklabel style = {font=\sffamily\footnotesize},
                    title style = {font=\scriptsize\sffamily},
                    ylabel near ticks,
                    y label style={font=\sffamily\footnotesize},
                    xlabel near ticks,
                    x label style={font=\sffamily\footnotesize},
                    legend cell align={left},
                    legend style={draw=none, font=\sffamily\footnotesize},
                    },
                    legend image code/.code={
                    \draw[mark repeat=2,mark phase=2]
                        plot coordinates {
                        (0cm,0cm)
                        (0.15cm,0cm)        
                        (0.3cm,0cm)         
                        };%
                    }
                    }
\pgfplotsset{compat=newest}

\begin{document}
\title{Entity Graphs for Exploring Online Discourse}
\author[1]{Nicholas Botzer}
\author[1,*]{Tim Weninger}

\affil[1]{University of Notre Dame, Notre Dame, IN}
\affil[*]{Corresponding author: tweninger@nd.edu}

\maketitle

\begin{abstract}
Vast amounts of human communication occurs online. These digital traces of natural human communication along with recent advances in natural language processing technology provide for computational analysis of these discussions. In the study of social networks the typical perspective is to view users as nodes and concepts as flowing through and among the user-nodes within the social network. In the present work we take the opposite perspective: we extract and organize massive amounts of group discussion into a concept space we call an \emph{entity graph} where concepts and entities are static and human communicators move about the concept space via their conversations. Framed by this perspective we performed several experiments and comparative analysis on large volumes of online discourse from Reddit. In quantitative experiments, we found that discourse was difficult to predict, especially as the conversation carried on. We also developed an interactive tool to visually inspect \emph{conversation trails} over the entity graph; although they were difficult to predict, we found that conversations, in general, tended to diverge to a vast swath of topics initially, but then tended to converge to simple and popular concepts as the conversation progressed. An application of the spreading activation function from the field of cognitive psychology also provided compelling visual narratives from the data. 
\end{abstract} 

\section{Introduction}

In any conversation, members continuously track the topics and concepts that are being discussed. The colloquialism ``train-of-thought'' is often used to describe the path that a discussion takes, where a conversation may ``derail,'' or ``come-full-circle,'' etc. An interesting untapped perspective of these ideas exists within the realm of the Web and Social Media, where a train-of-thought could be analogous to a trail over a graph of concepts. With this perspective, an individual’s ideas as expressed through language can be mapped to explicit entities or concepts, and, therefore, a single argument or train-of-thought can be treated as a path over the graph of concepts. Within a group discussion, the entities, concepts, arguments, and stories can be expressed as a set of distinct paths over a shared concept space, what we call an \textit{entity graph}.

Scholars have long studied discourse and the flow of narrative in group conversations, especially in relation to debates around social media~\cite{page2015narrative} and intelligence \cite{mateas2003narrative}. The study of language and discourse is rooted in psychology \cite{chafe2017language} and consciousness \cite{chafe1994discourse}.

%
Indeed, the linguist Wallace Chafe considered ``...conversation as a way separate minds are connected into networks of other minds.'' \cite{chafe2017language}
Looking at online conversations from this angle, a natural hypothesis arises: If we think of group discussion as a graph of interconnected ideas, then can we learn patterns that are descriptive and predictive of the discussion? 

Fortunately, recent developments in natural language processing, graph mining, and the analysis of discourse now permit the algorithmic modelling of human discussion in interesting ways by piecing them together. This is a broad goal, but in the present work we provide a first step towards graph mining over human discourse. 

Another outcome of the digital age is that much of human discourse has shifted to online social systems. Interpersonal communication is now observable at a massive scale. Digital traces of emails, chat rooms, Twitter or other threaded conversations that approximate in person communication are commonly available. A newer form of digital group discussion can be seen in the dynamics of Internet fora where individuals (usually strangers) discuss and debate a myriad of issues.

Technology that can parse and extract information from these conversations currently exists and operates with reasonable accuracy. From this large body of work, the study of \emph{entity linking} has emerged as a way to ground conversational statements to well-defined entities, such as those that constitute knowledge bases and knowledge graphs~\cite{shen2014entity}. Wikification, \textit{i.e.}, where entities in prose are linked to Wikipedia-entries as if it was written for Wikipedia, is one example of entity linking~\cite{cheng2013relational}. The Information Cartography project is another example that uses these NLP-tools to create visualizations that help users understand how related news stories are connected in a simple, yet meaningful manner \cite{shahaf2013information, shahaf2010connecting, keith2021narrative}. But because entity linking techniques have been typically trained from Wikipedia or long-form Web-text, they have a difficult time accurately processing conversational narratives, especially from social media~\cite{derczynski2015analysis}. Fortunately, recent progress in \emph{Social-NLP} has made considerable strides in recent years~\cite{ran2018attention} providing the ability to extract grounded information from informal, threaded online discourse~\cite{kolitsas2018end}.

\begin{figure}[t]
    \centering
    \includegraphics[width=.7\textwidth]{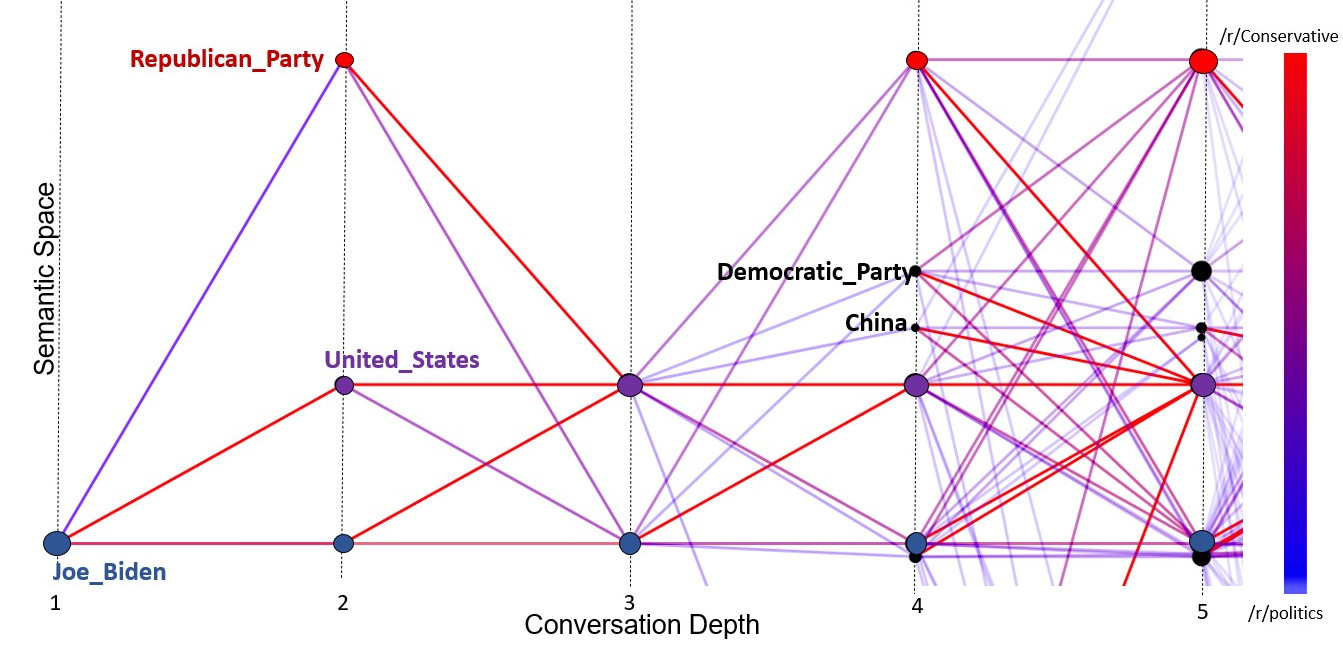}
    \caption{ Illustration of an entity graph created from threaded conversations from r/politics (blue-edges) and r/conservative (red-edges). The x-axis represents the (threaded) depth at which each entity was mentioned within conversations, extracted from Reddit, rooted at \textsf{Joe\_Biden}. The y-axis represents the semantic space of each entity, {\textit i.e.}, similar entities are closer than dissimilar entities on the y-axis. Edge colors denote whether the transition from one entity set to another occurs more often from one groups conversations than another. Node colors represent equivalent entity sets along the x-axis. In this visualization we observe a pattern of affective polarization as comments coming from /r/Conservative are more likely to drive the conversation towards topics related to the opposing political party.}
    \label{fig:con_lib}
\end{figure}

Taking this perspective, the present work studies and explores the flow of entities in online discourse through the lens of \emph{entity graphs}. We focus our attention on discussion threads from Reddit, but these techniques should generalize to online discussions on similar platforms so long as the entity linking system can accurately link the text to the correct entities. The threaded conversations provide a clear indication of the reply-pattern, which allows us to chart and visualize conversation-paths over entities.

To be clear, this perspective is the opposite of the conventional social networks approach, where information and ideas traverse over user-nodes; on the contrary, we consider discourse to be humans traversing over a graph of entities. The conventional approach to social networks is important for areas such as influence maximization \cite{kempe2003maximizing} and the spread of behaviors \cite{centola2010spread}. Instead, the goal of our alternative perspective is to discover this network of minds and uncover patterns of how they think over topics. This alternative perspective is motivated by the large number of influence campaigns \cite{schia2020hacking}, information operations \cite{weedon2017information}, and the effectiveness of disinformation \cite{glenski2019multilingual}. These campaigns often operate by seeding conversations in order to exploit conversation patterns and incite a particular group. Another motivation for our proposed methodology is humans attraction towards homophily and the large number of echo chambers that have been created online \cite{cinelli2021echo, garimella2018political}. Prior works \cite{garimella2018political} looking at echo chambers in political discourse rely on this notion of the ideas spreading between user-nodes. Other works looking at morality \cite{brady2020mad} also follow this notion of how moral text spreads throughout a user network. We stress here that our entity graph will allow for a flipped perspective of having users move across the graph of entities in various types of conversations. This position allows for a different form of analysis into how different groups or communities think as a whole.

Our way of thinking is illustrated in Fig.~\ref{fig:con_lib}, which shows a subset of path traversals, which we describe in detail later, from thousands of conversations in /r/politics and /r/conservative that start from the entity \textsf{Joe\_Biden}. As a brief preview, we find that conversations starting with \textsf{Joe\_Biden} tend to lead towards \textsf{United\_States} in conversations from the /r/conservative subreddit (indicated by a red edge), but commonly lead towards mentions of the \textsf{Republican\_Party} in conversations from /r/politics (indicated by blue-purple edge). From there the conversations move onward to various other entities and topics that are cropped from Fig~\ref{fig:con_lib} to maintain clarity. 

In the present work, we describe how to create entity graphs and use them to answer questions about the nature of online, threaded discourse. Specifically, we ask three research questions:

\begin{enumerate}
    \item[RQ1] How predictable is online discourse? Can we accurately determine where a conversation will lead?
    \item[RQ2] What do entity graphs of Reddit look like? In general, does online discourse tend to splinter, narrow, or coalesce? Do conversations tend to deviate or stay on topic?
    \item[RQ3] Can cognitive-psychological theories on spreading activation be applied to further illuminate and compare online discourse?
\end{enumerate}

We find that entity graphs provide a detailed yet holistic illustration of online discourse in aggregate that allow us to address our proposed research questions. Conversations have an enormous, visually random, empirical possibility space, but attention tends to coalesce towards a handful of common topics as the depth increases. Prediction is difficult, and gets more difficult the longer a conversation goes on. Finally, we show that entity graphs present a particularly compelling tool by which to perform comparative analysis. For example, we find, especially in recent years, that conservatives and liberals both tend to focus their conversations on the out-group -- a notion known as \emph{affective polarization}~\cite{iyengar2019origins}. We also find that users also tend to stick to the enforced topics of a subreddit as shown by how r/news tends towards entities from the United States and r/worldnews tends towards non-US topics.

\section{Methodology}
\subsection{Online Discourse Dataset}
Of all the possible choices from which to collect online discourse, we find that Reddit provides exactly the kind of data that can be used for this task. It is freely and abundantly available~\cite{baumgartner2020pushshift}, and it has a large number of users and a variety of topics. Reddit has become a central source of data for many different works~\cite{medvedev2017anatomy}. For example, recent studies on the linguistic analysis of Schizophrenia \cite{zomick2019linguistic}, hate speech~\cite{chandrasekharan2017you}, misogyny \cite{farrell2019exploring}, and detecting depression related posts \cite{tadesse2019detection} all make substantial use of Reddit data. 

The threading system that is built into Reddit comment-pages is important for our analysis. Each comment thread begins with a high level topic (the post title), that is often viewed as the start of a conversation around a specific topic. 
Users often respond to the post with their own comments. These can be viewed as direct responses to the initial post, and then each of these comments can have replies. This threading system generates a large tree structure where the root is the post title. Of course, such a threading system is only one possible realization of digital discussion, but this system provides the ability to understand how conversations move as users respond to each other in turn. Twitter, Facebook, and Youtube also have discussion sections, but it is very difficult to untangle who is replying to whom in these (mostly) unthreaded systems.

Reddit contains a large variety of subreddits, which are small communities focused on a specific topic. We limit our analysis to only a small number of them, but for each selection we obtain their complete comment history from January 2017 to June 2021. In total we selected five subreddits: /r/news, /r/worldnews, /r/Conservative, /r/Coronavirus and /r/politics. We selected these subreddits because they are large and attract a lot of discussion related to current events, albeit with their own perspectives and guidelines. These subreddits also contain a large number of entities, which we plan to extract and analyze.

Like most social sites, Reddit post-engagement follows the 90-9-1 rule of Internet engagement. Simply put, most users don't post or comment, and most posts receive almost no attention~\cite{medvedev2017anatomy}. 
Because of this we limit our data to include only those threads that are in the top 20\% in terms of number of comments per post. Doing so ensures that we mostly collect larger discussions threads that have an established back and forth. We also ignore posts from the well-known bot accounts, ({\emph e.g.}, AutoMod, LocationBot) to ensure we get actual user posts in the conversation.

\subsection{Entity Linking}

\begin{figure}[t]
\centering    
    \includegraphics[width=.90\textwidth]{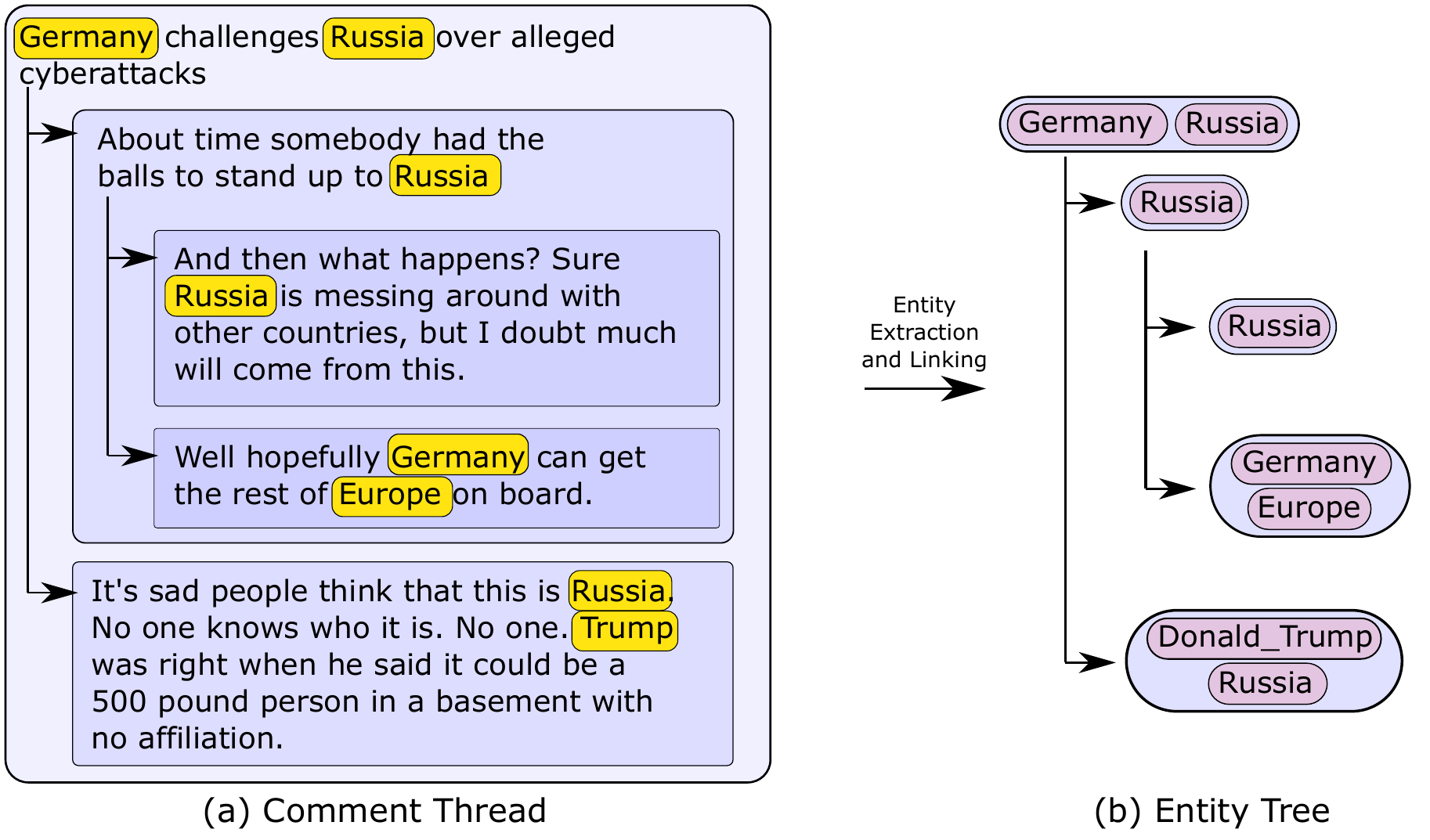}
 \caption{(Left) Example comment thread with the post title as the root, two immediate child comments, one of which has two additional child comments. Entity mentions are highlighted in yellow. (Right) The resulting entity tree where each comment is replaced by their entity set. Note the case where the mention-text Trump in the comment thread is represented by the standardized entity-label \textsf{Donald\_Trump} in the entity tree.}
 \label{fig:el_example}
\end{figure}

We use entity linking tools to extract the entities from each post title and comment in the dataset (\emph{c.f.} \cite{shen2014entity}).  Entity linking tools seek to determine parts of free form text that represent an entity (a mention) and then map that mention to the appropriate entity-listing in a knowledge base (disambiguation), such as Wikipedia. Existing models and algorithms rely heavily on character matching between the mention-text and the entity label, but more-recent models have employed deep representation learning to make this task more robust~\cite{sevgili2020neural}.

An example of entity linking on a comment thread is illustrated in Fig.~\ref{fig:el_example}. Each comment thread $T$ contains a post $R$ which serves as the root of the tree $c_r\in T$ and comments $c_x\in T$, where subscript $r$ and $x$ serve to index the post title and a specific comment. Each comment can reply to the root $c\rightarrow r$ or to another comment $c_x\rightarrow c_y$ thereby determining a comment's depth $\ell\in [0,\ldots,L]$. Comments and post titles may or may not contain one or more entities $S(c)$. These entity sets are likewise threaded, such that $S(c_x)\rightarrow S(c_y)$ means that the entities in $c_x$ were responded to with the entities in $c_y$, \emph{i.e.}, $c_x$ is the parent of $c_y$. With this formalism, the entity linking task transforms a comment threads into an \emph{entity tree} as seen in Fig.~\ref{fig:el_example}.

Specifically, we utilize the End-to-End (E2E) neural model created by Kolitaskas et. al. \cite{kolitsas2018end} to perform entity linking on our selected subreddits. Previous work has shown that entity linking on Reddit can be quite challenging due to the wide variety of mentions used \cite{botzer2021reddit}. The E2E model we use has been shown to have a high level of precision on Reddit but lacks a high recall \cite{botzer2021reddit}. We find using this model appropriate as we want to ensure that the entities we find are correct and reliable, but acknowledge that it may miss a portion of the less well-known entities, as well as missing any new entities that arise from entity drift. The choice of this entity linker also influenced our decision to analyze the selected subreddits as the performance is better in these selected subreddits. We also experimented with the popular REL entity linker \cite{van2020rel}. Although it did retrieve many more entities from the comments, we found a large number of the entities to be incorrect.




\begin{table}[t]\footnotesize
\centering
\caption{Reddit discourse dataset. Top 20\% of posts in terms of number of comments from five subreddits between January 2017 to June 2021.}

\begin{tabular}{@{}lrrrr@{}}
    \toprule 
  & \textbf{\# Posts} & \textbf{\# Comments} &  \textbf{Total Entities} &  \textbf{Unique Entities} \\ \midrule
/r/news  & 7,299  & 106,428 &  240,009 & 10,573  \\ 
/r/worldnews & 16,056  & 263,227 &  692,735 & 12,840 \\ 
/r/politics          & 15,596  & 326,958 &  756,576 & 11,908 \\ 
/r/Conservative          & 3,093   & 41,439  &  100,756 & 4,308  \\ 
/r/Coronavirus          & 18,469   & 252,303  &  509,632 & 10,246 \\ 
\end{tabular}

\label{tab:subreddit_stats}
\end{table}

Using the E2E model we extract entities from each post title and comment individually and construct the entity tree as illustrated in Fig~\ref{fig:el_example}. Table \ref{tab:subreddit_stats} shows a breakdown of the post, comment, and entity statistics for each subreddit considered in the present work.

\subsection{Entity Graph}

Given an entity tree, our next task is to construct a model that can be used to make predictions about the shape and future of the conversation, but also can be used as a visual, exploratory tool. Although entity trees may provide a good picture for a single conversation, we want to investigate patterns in a broader manner. To do this we consider conversations coming from a large number of entity trees in aggregate. This model takes the form of a weighted directed graph $G=(V,E, w)$ where each vertex $v \in V$ is a tuple of an entity set $S(c)$ and it's associated depth $\ell$ in the comment tree $v = (S(c), \ell)$. Each directed edge in the graph $e \in E$ connects two vertices $e = (v_1, v_2)$ such that the depth , $\ell$ of $v_1$ must be one less than the depth of $v_2$. Each edge in the graph $e \in E$ also contains a weight $w:E \rightarrow \mathbb{R}$ that represents the frequency of the transition from one entity set to another.
This directed graph captures not only the specific concepts and ideas mentioned within the discourse, but also the conversational flow over those concepts.

\begin{figure}
\centering    
    \includegraphics[width=.99\textwidth]{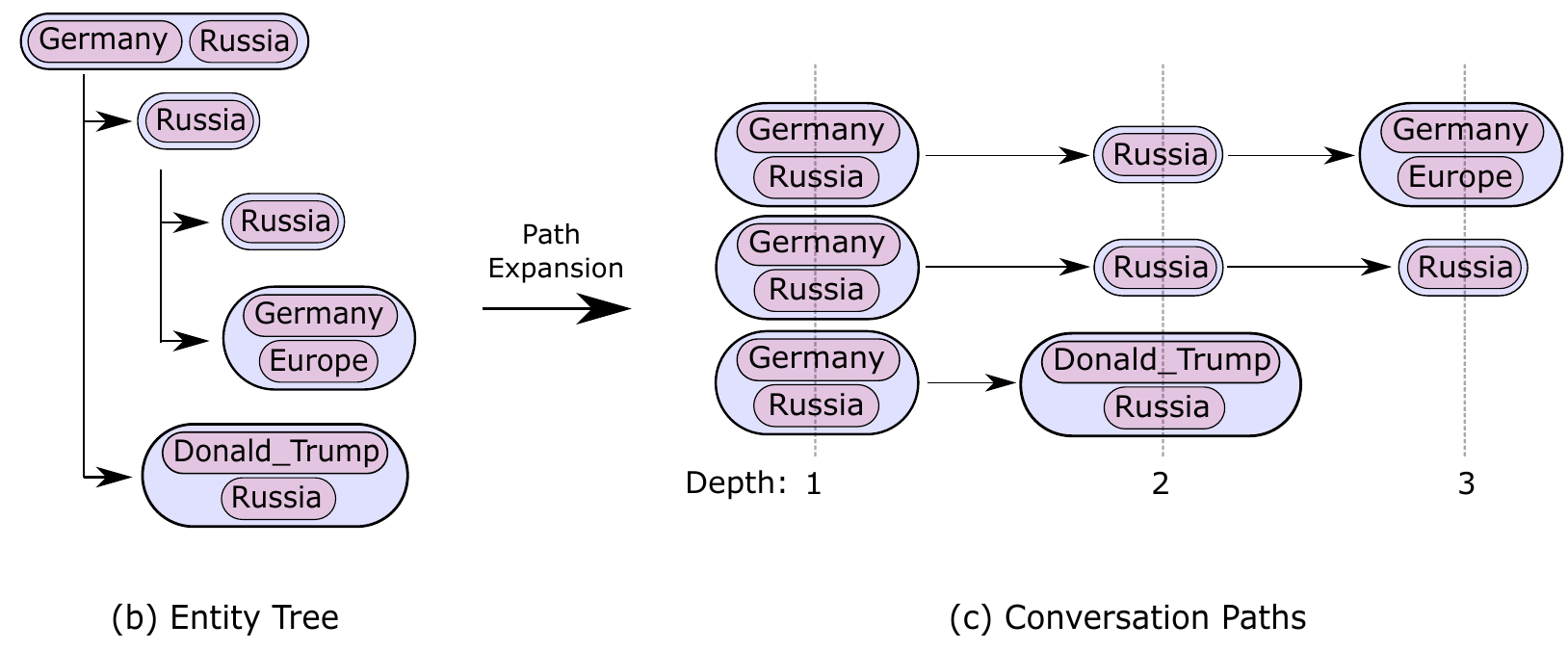}
 \caption{Paths extracted from the entity tree in Fig.~\ref{fig:el_example}(b) represented by directed edges over entity sets.}
 \label{fig:entitypath}
 \end{figure}

Continuing the example from above, Fig.~\ref{fig:entitypath} shows three individual paths $P$ representing the entity tree from Fig.~\ref{fig:el_example}(b). Each entity set moves from one depth to the next, representing the progression of the discussion.

During the construction of the entity paths, we remove comments that do not have any replies. Short paths, those with a length less than three, do not offer much information in terms of how the conversation will progress, because the conversation empirically did not progress. It may be useful to analyze why some topics resulted in no follow-on correspondence, but we leave this as a matter for future work.

Because we wish to explore online discourse in aggregate, this is the point where we aggregate across many comment threads $T\in\mathcal{T}$ where $\mathcal{T}$ represents an entire subreddit or an intentional mixture of subreddits depending on the task. We extract all of the conversation paths from our comment threads $\mathcal{T}$ to now have a group of conversation paths $\mathcal{P}$. To generate our graph we iterate over our group of paths $\mathcal{P}$ and aggregated them together to construct our entity graph. For every instance of an entity set transition in a conversation path we increment the weight $w$ of it's respective edge in our entity graph. One key aspect of this is that we count this transition only once per each comment thread $T$. This ensures that entity transitions do not get over counted, by virtue of the thread being larger and containing more conversation paths overall. 

One of the limitations of the current graph structure is that the graph does not capture conversation similarities if some of the entities overlap between two different vertices. For instance, another entity tree may result in having an entity set $S(c_r)$ that contains a subset of the entities in a given vertex. This new entity set may have a similar conversational flow but will not be captured in our current entity graph because the model does not allow for any entity overlap.



To help alleviate this issue we borrow from the notion of a hypergraph and perform a star-expansion on our graph $G$ \cite{zien1999multilevel}.
A hypergraph is defined as $\mathcal{H} = (X, E)$ where $X$ is the set of vertices and $E$ is a set of non-empty subsets of $X$ called hyperedges.
The star expansion process turns a hypergraph into a simple, bipartite graph. It works by generating a new vertex in the graph for each hyperedge present in the hypergraph and then connects each vertex to each new hyperedge-vertex. 
This generates a new graph $G(V,E)$ from $\mathcal{H}$ by introducing a new vertex and edge for each hyperedge such that $V = \mathcal{E}\cup\mathcal{P}$.

\begin{figure}
\centering    
    \includegraphics[width=.95\textwidth]{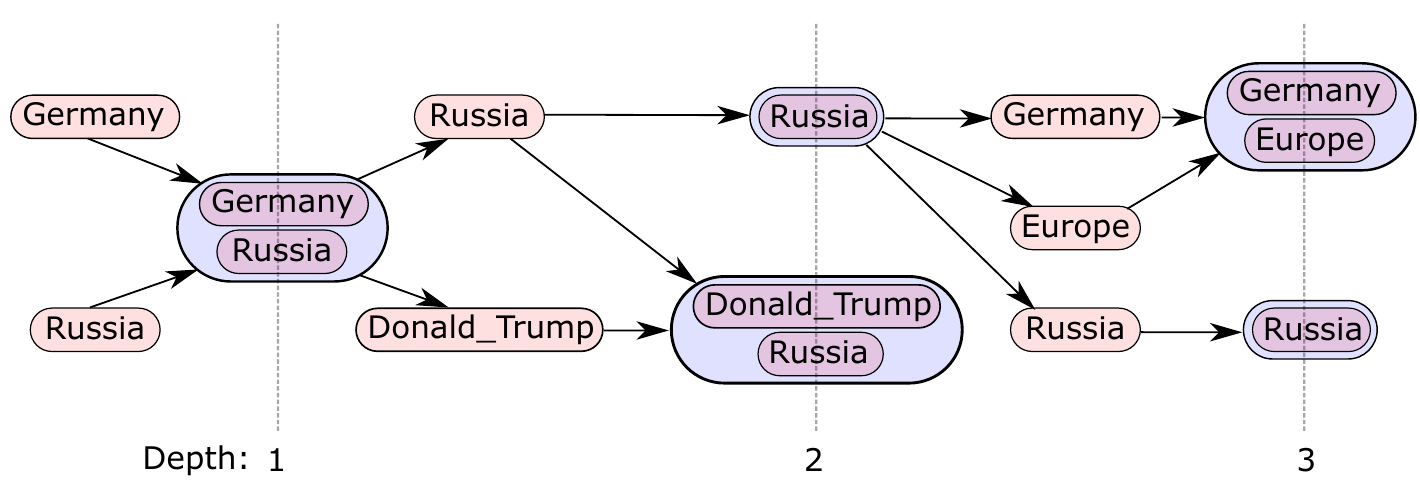}
 \caption{Entity graph constructed from a star-expansion of the entity tree in Fig~\ref{fig:el_example}(b) and the conversation paths in Fig.~\ref{fig:entitypath}(c) This model represents the entities, their frequent combinations, and the paths frequently used in their invocation. }
 \label{fig:narrative_hypergraph} 
 \end{figure}

While our model is a graph we can treat each entity set $S(c)$ as a hyperedge in our case to perform this star expansion. This will give us new vertices to represent each individual entity and allow us to capture transitions from one entity set to another if they share a subset of entities. An example of the resulting graph after performing a star-expansion can be seen in Fig.~\ref{fig:narrative_hypergraph}. This helps to provide valid transition paths that would otherwise not exist without the star expansion. When the star expansion operation is performed the edge weights between the new individual entity vertices and their respective entity sets is set to the number of times that entity set occurred at a given depth $l$. Although the star expansion process will generate a much larger graph due to the large number of vertices, it proves to be useful for prediction and aligning entity set vertices in a visual space.

This graph-model therefore represents the entities, their frequent combinations, and the paths frequently used in their invocation over a set of threaded conversations.

\section{Conversation Prediction}
Having generated these entity graphs we turn our attention to the three research questions. RQ1 first asks if these entity graphs can be used to predict where a conversation may lead. Clearly this is a difficult task, but recent advances in deep learning and language models have led to major improvements and interest in conversational AI~\cite{ram2018conversational}, which has further lead to the development of a number of models that utilize entities and knowledge graphs~\cite{yu2020survey} from various sources including Reddit~\cite{zhang2019grounded}. The main motivation of these tools is to use the topological structure of the knowledge graphs (entities and their relationships) to improve a conversational agents' ability to more-naturally select the next entity in the conversation. The typical methodology in related machine learning papers seeks to predict the next entity in some conversation~\cite{moon2019opendialkg}. In these cases, a dataset of paths through a knowledge graph is constructed from actual human conversations as well as one or more AI models. Then a human annotator picks the entity that they feel is most natural~\cite{moon2019opendialkg,jung2020attnio}.

Our methodology varies from these as we are not focused on making a machine learning model to accurately predict these entities precisely. Our goal is to demonstrate more broad patterns of people conversing over and through the topics. To this end, we do not evaluate with a standard machine learning paradigm aiming to optimize for metrics such as accuracy, precision, recall, etc. 
 To demonstrate that our entity graph captures broad patterns that can be further explored we perform two tasks: (1) the generalization task and (2) a similarity prediction task.  Each task uses 5-fold cross validation where we split the entity graph into 80/20 splits for $\mathcal{H}_{\textrm{train}}$ and $\mathcal{H}_{\textrm{test}}$ respectively. We perform this cross validation in a disjoint manner with the Reddit threads that we have extracted. This creates 5 different entity graphs, one for each split, and validates the model's generalization to unseen Reddit threads. Although this disjoint split ensures the threads are separate, we do not consider the temporal aspect of these threads.
 

\begin{figure}
   \centering
    \begin{tikzpicture}
\pgfplotstableread{

depth   percent y_error

1  0.917  0.004
2  0.954  0.004
3  0.665  0.007
4  0.611  0.014
5  0.574  0.013
6  0.569  0.016
7  0.519  0.015
8  0.618  0.025

}{\news}

\pgfplotstableread{
    depth   percent y_error

1  0.984  0.001
2  0.985  0.001
3  0.733  0.004
4  0.743  0.002
5  0.72  0.003
6  0.679  0.004
7  0.631  0.017
8  0.608  0.011

}{\worldnews}

\pgfplotstableread{
 depth percent y_error

1  0.979  0.001
2  0.986  0.001
3  0.754  0.003
4  0.713  0.002
5  0.653  0.012
6  0.659  0.009
7  0.661  0.011
8  0.604  0.011

}{\politics}

\pgfplotstableread{
 depth percent y_error
1  0.964  0.003
2  0.966  0.005
3  0.691  0.008
4  0.679  0.012
5  0.654  0.012
6  0.601  0.002
7  0.587  0.035
8  0.691  0.053

}{\conservative}

\pgfplotstableread{
 depth percent y_error
1  0.983  0.002
2  0.986  0.001
3  0.778  0.002
4  0.759  0.006
5  0.708  0.008
6  0.664  0.008
7  0.709  0.011
8  0.671  0.013

}{\coronavirus}

\begin{axis}[xlabel=Conversation Depth ($\ell$),
            ylabel=Percentage of Valid Entities,
            legend pos=north east,
            legend cell align=left,
            width=2.75in
            ]
  
  \addplot+[error bars/.cd,y dir=both,y explicit] 
    table[x=depth,y=percent,y error=y_error] {\news}; \addlegendentry{/r/news}
    \addplot+[error bars/.cd,y dir=both,y explicit] 
    table[x=depth,y=percent,y error=y_error] {\worldnews}; \addlegendentry{/r/worldnews}
    \addplot+[error bars/.cd,y dir=both,y explicit] 
    table[x=depth,y=percent,y error=y_error] {\politics}; \addlegendentry{/r/politics}
    \addplot+[error bars/.cd,y dir=both,y explicit] 
    table[x=depth,y=percent,y error=y_error] {\coronavirus}; \addlegendentry{/r/coronavirus}
    \addplot+[error bars/.cd,y dir=both,y explicit] 
    table[x=depth,y=percent,y error=y_error] {\conservative}; \addlegendentry{/r/conservative}
    
\end{axis}
\end{tikzpicture}
    \caption{Percent of the predictions made on the testing set that, on average, exist in the training set for 5-folds. Higher is better.}
    \label{fig:valid_predictions}
\end{figure}
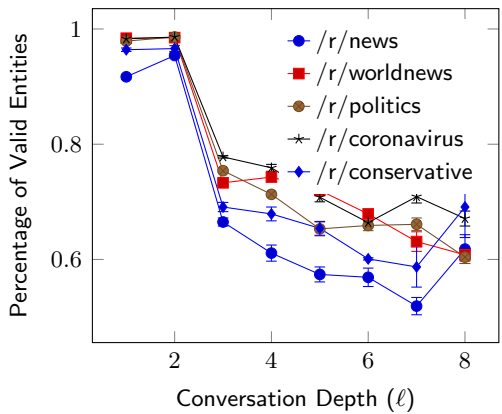

The first task: generalization, gets at the heart of our broader question on the predictability of conversation paths. In this task we simply calculate the number of entity sets, at each level  in $\mathcal{H}_{\textrm{test}}$ that also appear in the same level in $\mathcal{H}_{\textrm{train}}$ of our entity graph. Formally, we measure generalization as $1-\frac{\|S_{\ell}\in \mathcal{H}_{\textrm{test}}\setminus S_{\ell}\in \mathcal{H}_{\textrm{train}}\|}{\|S_{\ell}\in \mathcal{H}_{\textrm{test}}\|}$ for each $\ell$.

In simple terms, generalization tells us, given an unseen conversation comment, if the model can make a prediction from the given comment by matching at least one entity in our entity graph model. This task therefore validates how well the model captures general conversation patterns by matching at the entity level.
Results of this analysis are shown in Fig.~\ref{fig:valid_predictions} where color and shape combinations indicate the subreddit and $\ell$ is represented along the x-axis. Error bars represent the 95\% confidence interval of the mean across the 5 folds. We find that the entity graph captures much more of the information early in conversations. As the depth increases to three and beyond, we note a sharp drop in the overlap between the test and training sets. The widening confidence interval also indicates that the amount of information varies based on the test set. From these results, we conclude that analyzing the flow of an unseen conversation early-on is reasonable, but findings from deeper in the conversation may be difficult because key entities may be missing from the entity graph.

The second task: similarity prediction looks to measure the similarity between a predicted entity set and the actual entity set. This methodology uses the entity embeddings from the E2E entity linking model to represent the entities in the vector space. For each root in $\mathcal{H}_{\textrm{test}}$ we find its matching root in the $\mathcal{H}_{\textrm{train}}$; if a match does not exist, we discard and start again. Then we make the Markovian assumption and perform probabilistic prediction for each path in the training set via $Pr(S_{\ell+1}(c_y) \vert S_{\ell}(c_x))$, \emph{i.e.}, the empirical probability of a conversation moving to $S_{\ell+1}(c_y)$ given the conversation is currently at $S_{\ell}(c_x)$ in $\mathcal{H}_{\textrm{train}}$. The probability for each transition is based on the edge weights that we captured during the graph construction step. As determined in the previous experiment, entity sets are increasingly unlikely to match exactly as the depth increases; so rather than a 0/1 loss, we measure the word movers distance (WMD) between the predicted entities and the actual entities~\cite{kusner2015word}.

\begin{figure}
    \centering
    \begin{tikzpicture}

\begin{groupplot}[boxplot/draw direction = y,
    group style={
        group size=3 by 1,
          horizontal sep=.2cm,
          vertical sep=.1cm,
                    ylabels at=edge left,
          yticklabels at=edge left,
            xlabels at=edge bottom,
      },
      xmin=0,
      ymin=-.3,
      ymax=6.3,
       xlabel=Conversation Depth ($\ell$),
      width = 2.00in]

\nextgroupplot[title=/r/news, ylabel=WMD]
\addplot+ [] plot coordinates {
    (1, 3.2791650737073565)
    (2, 3.2635300234556577)
    (3, 3.3276169830221196)
    (4, 3.3878750939343787)
    (5, 3.715213563218816)
    (6, 3.6632916518870497)
    (7, 3.8453914496255637)
    };
    
] coordinates {};

\addplot[boxplot prepared={
    median=3.2791650737073565,
    upper quartile=3.6459243938722574,
    lower quartile=2.8999729866021324,
    upper whisker=4.624422207487423,
    lower whisker=1.6992104524666407
}
] coordinates {};
\addplot[boxplot prepared={
    median=3.2635300234556577,
    upper quartile=3.6586357589671072,
    lower quartile=2.899441928346634,
    upper whisker=4.312804308767475,
    lower whisker=1.8250988184755916
}
] coordinates {};
\addplot[boxplot prepared={
    median=3.3276169830221196,
    upper quartile=3.7445162117884103,
    lower quartile=2.8476071595049683,
    upper whisker=4.336010707431907,
    lower whisker=0.0
}
] coordinates {};
\addplot[boxplot prepared={
    median=3.3878750939343787,
    upper quartile=3.7924903087404718,
    lower quartile=3.06130608019972,
    upper whisker=4.45761593127812,
    lower whisker=2.0868185707771496
}
] coordinates {};
\addplot[boxplot prepared={
    median=3.715213563218816,
    upper quartile=3.9780286009608403,
    lower quartile=3.295397158414559,
    upper whisker=4.358494988308137,
    lower whisker=2.187823402103025
}
] coordinates {};
\addplot[boxplot prepared={
    median=3.6632916518870497,
    upper quartile=3.9738517328451595,
    lower quartile=3.5214059727945384,
    upper whisker=4.233569397487223,
    lower whisker=2.590580140408539
}
] coordinates {};
\addplot[boxplot prepared={
    median=3.8453914496255637,
    upper quartile=3.9595342980123167,
    lower quartile=3.5394300993342114,
    upper whisker=4.519144923722868,
    lower whisker=3.2411538810563156
}
] coordinates {};

\nextgroupplot[title=/r/worldnews,]

\addplot+ [] plot coordinates {
    (1, 3.2030613824252487)
    (2, 3.234701298744714)
    (3, 3.290440226792721)
    (4, 3.2909211268410927)
    (5, 3.3277520818883244)
    (6, 3.5254879207738696)
    (7, 3.351430671651425)
    (8, 3.6341572526134653)
    };
    
] coordinates {};

\addplot[boxplot prepared={
    median=3.2030613824252487,
    upper quartile=3.606128428907953,
    lower quartile=2.8447747408211033,
    upper whisker=4.933577957373898,
    lower whisker=0.0
}
] coordinates {};
\addplot[boxplot prepared={
    median=3.234701298744714,
    upper quartile=3.712317197200945,
    lower quartile=2.8882434368180694,
    upper whisker=4.933577957373898,
    lower whisker=0.0
}
] coordinates {};
\addplot[boxplot prepared={
    median=3.290440226792721,
    upper quartile=3.80558525524633,
    lower quartile=2.8662449624060797,
    upper whisker=4.413715098626627,
    lower whisker=0.0
}
] coordinates {};
\addplot[boxplot prepared={
    median=3.2909211268410927,
    upper quartile=3.825494958229752,
    lower quartile=2.8790021420980896,
    upper whisker=4.6187811486477415,
    lower whisker=0.0
}
] coordinates {};
\addplot[boxplot prepared={
    median=3.3277520818883244,
    upper quartile=3.7971499980842744,
    lower quartile=2.9784739748296207,
    upper whisker=4.494600349329982,
    lower whisker=2.073021906105208
}
] coordinates {};
\addplot[boxplot prepared={
    median=3.5254879207738696,
    upper quartile=3.9519714374371007,
    lower quartile=3.061430881975947,
    upper whisker=4.391827120189745,
    lower whisker=2.0309917596323666
}
] coordinates {};
\addplot[boxplot prepared={
    median=3.351430671651425,
    upper quartile=3.729795505256313,
    lower quartile=3.0526838231536426,
    upper whisker=4.379344438325749,
    lower whisker=1.7714492764135803
}
] coordinates {};
\addplot[boxplot prepared={
    median=3.6341572526134653,
    upper quartile=3.96082194992269,
    lower quartile=3.1457025119100175,
    upper whisker=4.193215096744151,
    lower whisker=2.3864219495654195
}
] coordinates {};

\nextgroupplot[title=/r/politics]
\addplot+ [] plot coordinates {
    (1, 2.7973301792680068)
    (2, 2.9339564428522005)
    (3, 3.0479249996261575)
    (4, 3.0582878594301572)
    (5, 3.3335450516080005)
    (6, 3.5469824208693908)
    (7, 3.83179128960357)
    (8, 3.8708275954538394)
    };
    
] coordinates {};
\addplot[boxplot prepared={
    median=2.7973301792680068,
    upper quartile=3.2122979207594815,
    lower quartile=2.3835473384845693,
    upper whisker=4.305127790969139,
    lower whisker=0.0
}
] coordinates {};
\addplot[boxplot prepared={
    median=2.9339564428522005,
    upper quartile=3.451972870704283,
    lower quartile=2.507327927797877,
    upper whisker=4.380467273993364,
    lower whisker=0.0
}
] coordinates {};
\addplot[boxplot prepared={
    median=3.0479249996261575,
    upper quartile=3.6468625195922204,
    lower quartile=2.563127661241244,
    upper whisker=4.34603039608814,
    lower whisker=1.0574232309091243
}
] coordinates {};
\addplot[boxplot prepared={
    median=3.0582878594301572,
    upper quartile=3.6097621906960784,
    lower quartile=2.640194960903214,
    upper whisker=4.387937915350033,
    lower whisker=1.4732698134192097
}
] coordinates {};
\addplot[boxplot prepared={
    median=3.3335450516080005,
    upper quartile=3.8659188551922847,
    lower quartile=2.7553833241578083,
    upper whisker=4.229702577284236,
    lower whisker=1.5890362708103511
}
] coordinates {};
\addplot[boxplot prepared={
    median=3.5469824208693908,
    upper quartile=3.943226177076828,
    lower quartile=3.0116157520082387,
    upper whisker=4.485238703254283,
    lower whisker=1.7552439421585437
}
] coordinates {};
\addplot[boxplot prepared={
    median=3.83179128960357,
    upper quartile=3.960437936628809,
    lower quartile=3.031641698876413,
    upper whisker=4.0940525079054675,
    lower whisker=0.0
}
] coordinates {};
\addplot[boxplot prepared={
    median=3.8708275954538394,
    upper quartile=4.027313202125455,
    lower quartile=2.780513292578112,
    upper whisker=4.2187730200911835,
    lower whisker=2.2275030037768544
}
] coordinates {};

\end{groupplot}
\end{tikzpicture}
    \caption{Box plot of Word Movers Distance (WMD) as a function of the conversation depth $\ell$. Lower is better. Box plots represent WMD-error of entity representations predicted by the narrative hypergraph over all entities, over all depth, over five folds. }
    \label{fig:distance_box_plot}
\end{figure}
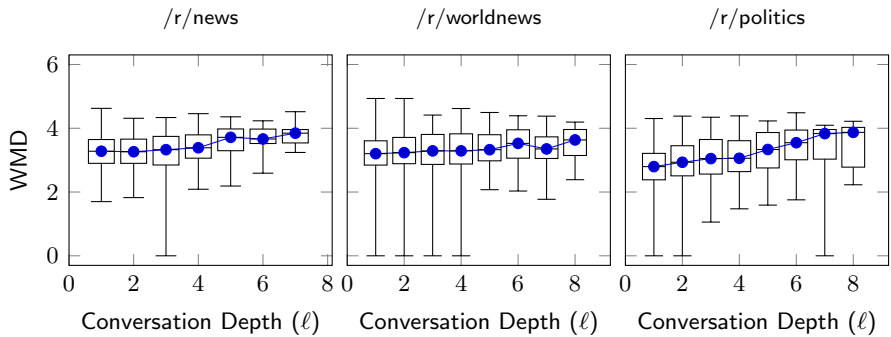

Results for this comparison are shown in Fig.~\ref{fig:distance_box_plot} for three of the larger subreddits. We again find that as the depth of the conversation increases the distance between our predicted tree and the ground truth entities rises. These results indicate that as a conversation continues, the variety of topics discussed tends to increase. Therefore, predictions are likely to not align well very to those of the true conversation. This is most clearly seen in the /r/politics plot in Fig.~\ref{fig:distance_box_plot}, where we note a sharp increase in the later parts of the conversation. If the variety of topics was consistent, then we would expect the WMD to stay relatively flat throughout the conversation depth.

\section{Conversation Traversals}
Next, we investigate RQ2 through a visualization of the entity graph. Recall that the entity graph contains entity sets over the depths of the conversation. Specifically, we seek to understand what conversations on Reddit look like. Do they splinter, narrow, or behave in some other way? We call the set of visual paths \emph{conversation traversals} because they indicate how users traverse the entity graph. 

We generate these visual conversation traversals using a slightly modified force directed layout~\cite{fruchterman1991graph}. Graph layout algorithms operate like graph embedding algorithms LINE, node2vec, etc, but rather than embedding graphs into a high dimension space, visual graph layout tools embed nodes and edges into a 2D space. In our setting we do make some restrictions to the algorithm in order to force topics to coalesce into a visually meaningful and standardized space.
Specifically, we fix the position of each vertex in our graph on the x-axis according to $\ell$. As in Fig.~\ref{fig:narrative_hypergraph}, individual entity vertices always occur to the left of entity set vertices, making the visualization illustrate how conversations flow from the start to finish in a left to right fashion. 

This restriction forces the embedding algorithm to adjust the position only on the y-coordinate, and this is necessary to allow the individual entity to entity set edges from the star-expansion to pull entity set vertices close together if and only if they share many common entities. Loosely connected or disconnected entities will therefore not be pulled together. As a result, the y-axis tends to cluster entities and entity-sets together in a semantically meaningful way.

Embedding algorithms are typically parameterized with a learning rate parameter that determines how much change can happen to the learned representation at each iteration. Because we want entities to be consistent horizontally, we modify the learning rate function to increasingly dampen embedding updates over 100 iterations per depth. For example, given a entity graph of depth $L=10$, we would expect 1,000 iterations total. We initially allow all entities and entity sets to update according to the default learning rate, but as the iterations increase to 100 the learning rate of the entities and entity sets at $\ell=1$ will slowly dampen and eventually lock into place at iteration 100.
When these entities and entity sets lock we also lock those same entities and entity sets at all other depths. This ensures that each of these entities and entity sets will be drawn as a horizontal line at the given y position.

Then, from iterations 100-200, the learning rate of the entities and entity sets at $\ell=2$ will slowly dampen and eventually lock into place at iteration 200. Meanwhile the entities and entity sets at deep levels will continue to be refined. In this way, the semantically meaningful y-coordinates tend to propagate from left to right as the node embedding algorithm iterates.

One complication is that the sheer number of entities and the conversation paths over the entities is too large to be meaningful to an observer. So we do not draw the entity-nodes generated by the star-expansion and instead opt to rewire entities sets based on the possible paths through the individual entity nodes. We also tune the edge opacity based on the edge weights. 

 
 \begin{figure}
    \centering
    \includegraphics[width=\textwidth]{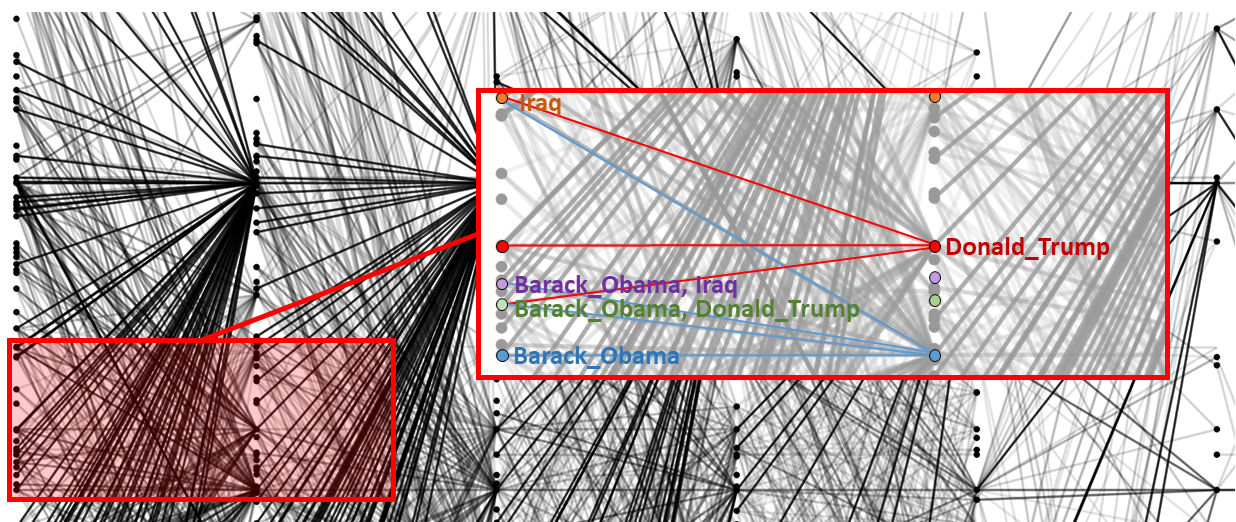}
    \caption{Entity graph showing the visual conversation traversals from /r/news. This illustration shows the paths of conversations over entity sets. The x-axis represents the depth of the conversation; entity sets are clustered into a semantically meaningful space along the y-axis. Inset graph highlights five example entity sets and their connecting conversation paths. Node colors represent equivalent entity sets. In this example we highlight how entity sets are placed in meaningful semantic positions in relation to one another.}
    \label{fig:news}
\end{figure}

We draw the resulting graph with D3 to provide an interactive visualization~\cite{bostock2011d3}. 

Conversation traversals of the entity graph generated from /r/news is illustrated in Fig.~\ref{fig:news}. This illustration is cropped to remove the four deepest vertical axes (on the right) and is also cropped to show the middle half of the illustration. A zoomed in version highlights some interesting entity sets present in the /r/news conversation. Recall that the entity sets are consistent horizontally so that both red circles on the left and the right of the inset plot both indicate the entity set with \textsf{Donald\_Trump}; likewise the blue circles on the left and the right of the insert both represent \textsf{Barack\_Obama}. Edges moving visually left to right indicate topical paths found in online discourse. In the /r/news subreddit, which tracks only US news, \textsf{Donald\_Trump} and \textsf{Barack\_Obama} are frequent visits, but so too are national entities like \textsf{United\_States} (not highlighted), \textsf{Iraq}, and others. It is difficult to see from this illustration, but the expanded interactive visualization shows a common coalescing pattern where large sets of entities and unique combinations of ideas typically coalesce into more simple singleton entities like \textsf{Barack\_Obama} or \textsf{United\_States}.

\begin{figure}
    \centering
    \includegraphics[width=.95\textwidth]{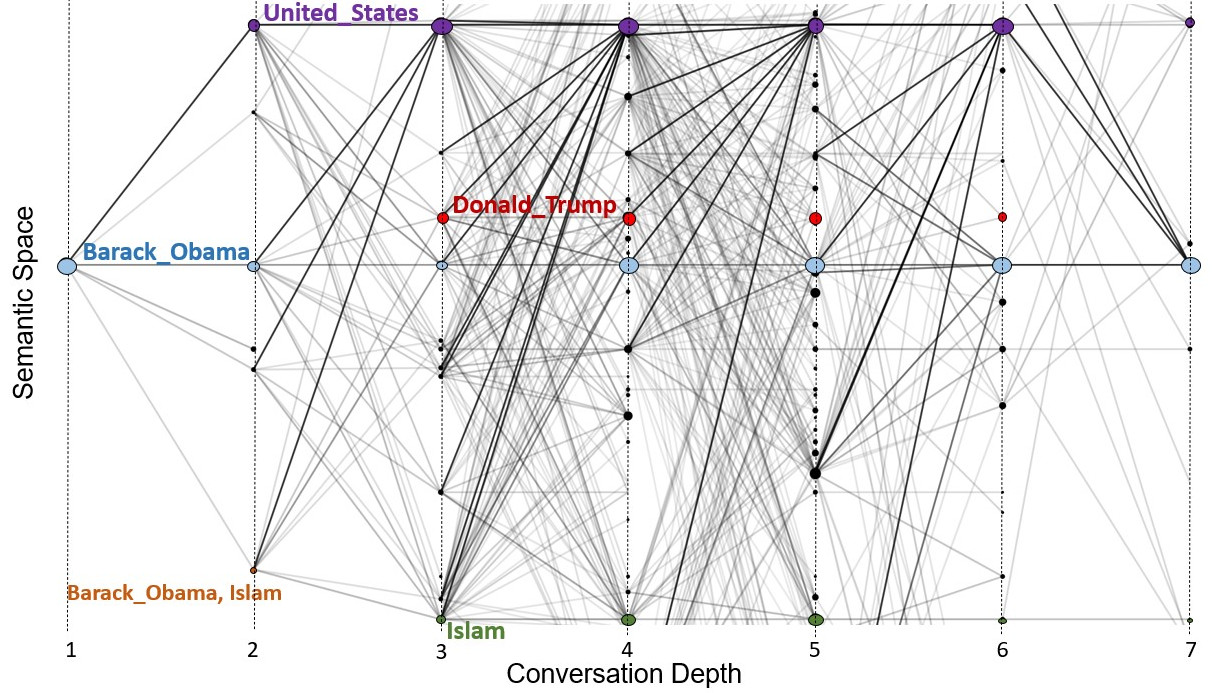}
    \caption{Entity graph example of spreading activation on /r/news when \textsf{Barack\_Obama} is selected as the starting entity. The x-axis represents the (threaded) depth at which each entity was mentioned within conversations rooted at \textsf{Barack\_Obama}. The y-axis represents the semantic space of each entity, {\textit i.e.}, similar entities are closer than dissimilar entities on the y-axis. Node colors represent equivalent entity sets. In this example, we observe that conversations starting from \textsf{Barack\_Obama} tend to center around the \textsf{United\_States}, political figures such as \textsf{Donald\_Tump}, and discussion around whether his religion is \textsf{Islam}.}
    \label{fig:sa_barack_obama}
\end{figure}

\subsection{Spreading Activation}
Next, we adapt the illustration of conversation traversals to begin to answer RQ3. Specifically, we are interested in how the differences in starting points, at the roots of the comment tree, have any impact on the eventual shape of the conversation. For example, given a conversation starting with \textsf{Donald\_Trump} how will the conversation take shape for liberals and how might that conversation be different among conservatives? This kind of analysis provides endless possibilities in the analysis of how different groups of people think and articulate ideas a given topic.

To help answer this question, we employ tools from the study of \emph{spreading activation}~\cite{collins1975spreading}. Spreading activation is a concept from cognitive psychology that has been used to model how ideas spread and propagate in the brain from an initial source. A popular use for spreading activation has been on semantic networks to find the relatedness between different concepts. 
Formally, spreading activation works by specifying two parameters: (1) a
firing threshold $F\in [0,\ldots,1]$ and (2) a decay factor $D\in [0,\ldots,1]$. The vertex/entity set selected by a user will be given an initial activation $A_i$ of 1. This is then propagated to each connected vertex as $A_i \times w_j \times D$ where $w_j$ is the weight of each edge connection to the corresponding vertex. Each vertex will then acquire its own activation value $A_i$ based on the total amount of signal received from all incoming edges. If a vertex has acquired enough activation to exceed the firing threshold $F$, it too will fire further propagating forward through the graph. In the common setting, vertices are only allowed to fire once and the spreading will end once there is no more vertices to activate.

In our work we use spreading activation as a method for a user to select a starting topic/entity set within the illustration of conversation traversals. The spreading activation function will then propagate the activation of entities along the conversation paths to highlight those that are mostly likely to activate from a given starting point. Because we permit the entity graph to be constructed (and labeled) from multiple subreddits, we can also use the spreading activation function to compare and contrast how users from different subreddits activate in response to a topic.

After spreading activation has been calculated, our interactive visualization tool removes all vertices and links that are not part of the activated portion of the graph. All of the vertices involved in spreading activation will have their size scaled based on how much activation they received. An example of this is cropped and illustrated in Fig.~\ref{fig:sa_barack_obama}, which shows how spreading activation occurs when the entity set \textsf{Barack\_Obama} is activated within /r/news. Here we see that conversations starting with (only) \textsf{Barack\_Obama} tend to move towards discussions about the United\_States. We also note that the \textsf{Islam} entity is semantically far away from \textsf{Barack\_Obama} and \textsf{Donald\_Trump} as indicated its placement on the y-axis. The results from using spreading activation allow for a much more granular investigation of conversational flow. These granular levels of conversational flow demonstrate that an individual can search for patterns related to influence campaigns, echo chambers and other social media maladies across a number of topics.

\section{Comparative Analysis}
The visual conversation traversals appears to be helpful for investigating trends within a group. But, our final goal is to use these to compare and contrast how different groups move through the conversation space. Our first attempt at this was to use and overlay separate plots and attempt to compare the trends. This would be challenging though because it would fail to capture the magnitude in any differences between the groups for various entity set transitions. Our second attempt, instead, modified the entity graph creation process to take in data from two different subreddits. By using both communities we can capture how often an entity transition occurs in each subreddit and use color gradients to indicate the relative strength of each transition probability based on the edge weight we find in each subreddit. This visually shows if correlations occur between subreddits. In the present work, we examined three different scenarios among the subreddits in our dataset.

\paragraph{Scenario 1: liberals and conservatives} Determining how motivated groups communicate about and respond to various topics is of enormous importance in modern communication studies. For example, communication specialists and political scientists are interested in understanding how users respond to coordinated influence campaigns that flood social media channels with the same message~\cite{paul2016russian}. Repetition is key for the idea to stick, and we would expect then that these forms of messaging would begin to appear in the  entity graphs and possibly visually indicated in the conversation traversals. 

Although a full analysis of this difficult topic is not within the purvue of the current work, we do perform a comparative analysis of /r/Conservative and /r/politics as proxies for comparing conservative and liberal groups, respectively. We pay particular attention to determining the particular topics and entities that each group tends to go towards later (deeper) in the conversation. Such a comparative analysis may be key to understanding how coordinated influence campaigns orient the conversation of certain groups or de-rail them. 

The comparative illustration using spreading activation was used at the beginning of the paper in Fig.~\ref{fig:con_lib} and is not re-illustrated in this section. The illustration yields some interesting findings. While one might expect /r/Conservative to discuss members or individuals related to the republican party, we instead find that conversations tend to migrate toward mentions of liberal politicians (\emph{e.g.}, \textsf{Joe\_Biden}) indicated by red lines in Fig.~\ref{fig:con_lib}. The reverse holds true as well: mentions of \textsf{Joe\_Biden} leads towards mentions of the \textsf{Republican\_Party} by the liberal group, as indicated by the blue line connecting the two. A brief inspection of the underlying comments reveals that users in each subreddit tend to talk in a negative manner towards the other party's politicians. This is a clear example of affective polarization \cite{iyengar2019origins} being captured by our visualization tool. Affective polarization is where individuals organize around principles of dislike and distrust towards the out-group (the other political party) even moreso than trust in their in-group. 

Another finding we observe is the more pronounced usage of the \textsf{United\_States} by conservatives than liberals. This observation could be explained by the finding that conservatives show a much larger degree of overt patriotism than liberal individuals~\cite{huddy2007american}, which has more recently lead to a renewed interest in populism and nationalism~\cite{de2017populism}.

\paragraph{Scenario 2: US news and Worldnews} In our second scenario, we compare the conversations from /r/news (red) and /r/worldnews (blue), which are geared towards US-only news and non-US news respectively.

\begin{figure}
    \centering
    \includegraphics[width=.95\textwidth]{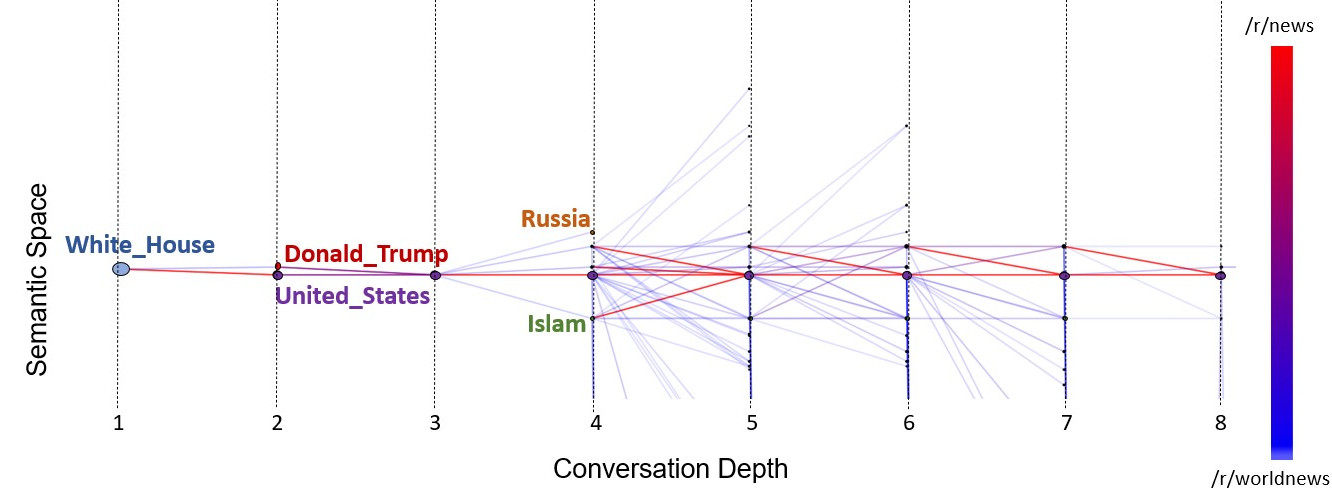}
    \caption{Illustration of an entity graph created from threaded conversations from /r/news (red-edges) and r/worldnews (blue-edges). The x-axis represents the (threaded) depth at which each entity set was mentioned within conversations rooted at \textsf{White\_House}. The y-axis represents the semantic space of each entity, {\textit i.e.}, similar entities are closer than dissimilar entity sets on the y-axis. Nodes colors represent equivalent entity sets. Conversations in /r/news tends to coalesce to \textsf{United\_States}, while conversations in /r/worldnews tend to scatter into various other countries (unlabeled black nodes connected by thin blue lines)}
    \label{fig:news_worldnews}
\end{figure}

The comparison between these subreddits reveals unsurprising findings. A much larger portion of the entity sets come from /r/worldnews as they discuss a much broader range of topics. Many of the entity transitions that are dominated by /r/worldnews come from discussions of other countries, events, and people outside of the United States. The aspects that are shown to come primarily from /r/news are topics surrounding the United States, China, and major political figures from the United States. An example of this can be seen in Fig.~\ref{fig:news_worldnews} which illustrates spreading activation starting from \textsf{White\_House}. Here, the dominating red lines, which reflects transitions from within conversations on /r/news, converge to \textsf{United\_States}, even after topics like \textsf{Russia} or \textsf{Islam} are discussed. An interesting side note is that many of the unlabeled entities entering the conversation via blue lines (/r/worldnews) in $\ell=5$ and $\ell=6$ represent other countries such as \textsf{Canada}, \textsf{Japan}, \textsf{Mexico}, and \textsf{Germany}. The findings from this comparative analysis do not show any extremely interesting results but, it does show that the entity graph is able to capture what one would see as the assumed patterns to find from comparing these two subreddits of interest.

\paragraph{Scenario 3: COVID and Vaccines} Our final analysis focuses on comparing a single subreddit, /r/Coronavirus, but during two different time periods. There is a large amount of work that has been done analyzing Covid online looking at partisanship \cite{rao2021political}, user reaction to misinformation \cite{kalantari2021characterizing}, and differences in geographic concerns \cite{guntuku2021twitter}. The first segment (highlighted in red) comes from the period of January through June in 2020, which was during the emergence of the novel Coronavirus. Although the /r/Coronavirus subreddit had existed for many years prior, it became extremely active during this time. The second segment was from the following year January - June 2021. This time period corresponded to the development, approval and early adoption of vaccines. 

\begin{figure}[t]
    \centering
    \includegraphics[width=.95\textwidth]{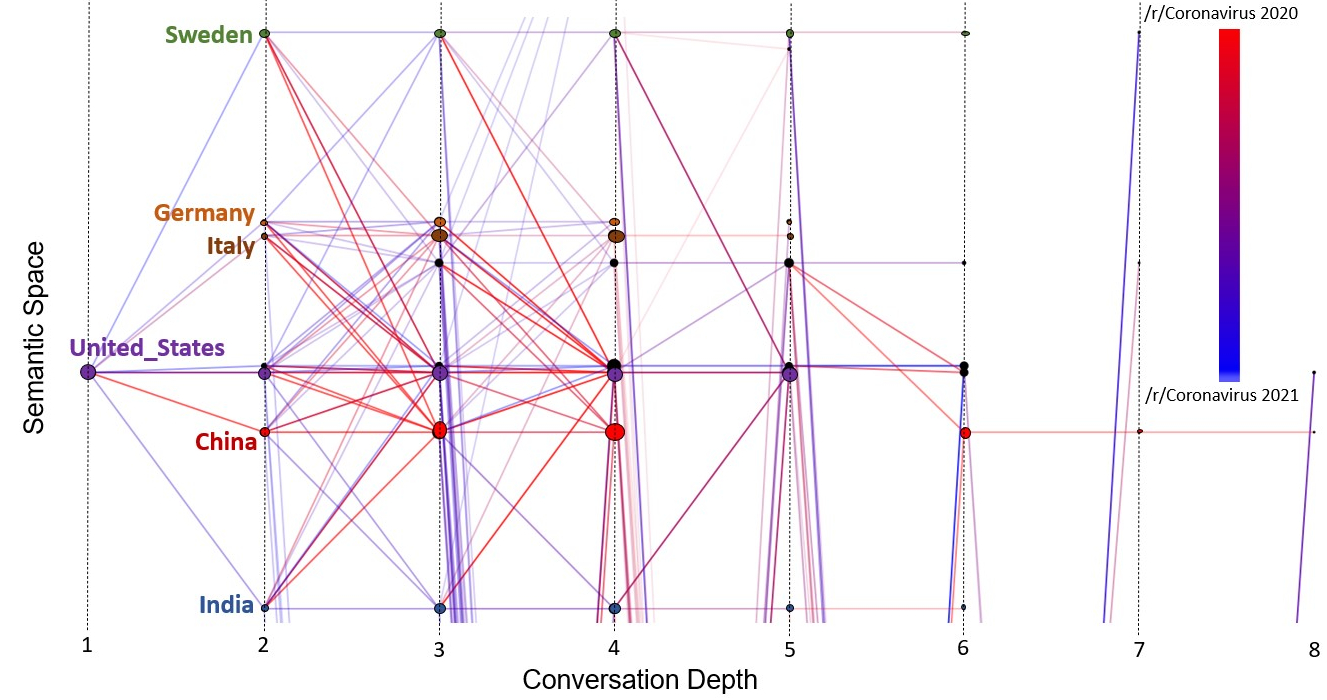}    \caption{Comparison between the first 6 months of /r/Coronavirus from 2020 to 2021. Illustration of an entity graph created from threaded conversations from /r/Coronavirus in Jan--June of 2020 (red-edges) and from Jan--June of 2021 (blue-edges). The x-axis represents the (threaded) depth at which each entity set was mentioned within conversations rooted at \textsf{United\_States}. The y-axis represents the semantic space of each entity set, {\textit i.e.}, similar entity sets are closer than dissimilar entity sets on the y-axis. Node colors represent equivalent entity sets. Conversations tended to focus on \textsf{Chin}a and \textsf{Italy} early in the pandemic, but turn towards a broader topic space later in the pandemic.}
    \label{fig:covid}
\end{figure}

Our analysis of this visualization yielded some interesting findings related to the coronavirus pandemic that we illustrate in Fig.~\ref{fig:covid}. 
If we begin spreading activation from the perspective of \textsf{United\_States} we find that most of the discussion leads to \textsf{China} and \textsf{Italy} in 2020, which appears reasonable because of China and Italy's early struggles with virus outbreaks. In comparison, the 2021 data appeared more likely to mention \textsf{Sweden}, \textsf{India}, and \textsf{Germany}, which had severe outbreaks during those months. Our findings from spreading activation allow us to capture the shifting changes in countries of interest from 2020 to 2021 as the pandemic progressed.

\section{Discussion}

In the current work we presented a new perspective by which to view and think about online discourse. Rather than taking the traditional social networks view where information flows over the human participants, our view is to consider human conversations as stepping over a graph of concepts and entities. We call these discourse maps \emph{entity graphs} and we show that they present a fundamentally different view of online human communication. 

Taking this perspective we set out to answer three research questions about (1) discourse prediction, (2) illustration, and (3) behavior comparisons between groups. We found that discourse remains difficult to predict, and this prediction gets harder the deeper into the conversation we attempt predictions. We demonstrate that the visual conversation traversals provide a view of group discourse, and we find that online discourse tends to coalesce into narrow, simple topics as the conversation deepens -- although those topics could be wildly different from starting topic. Finally, we show that the spreading activation function is able to focus the visualization to provide a comparative analysis of competing group dynamics. 

\subsection{Limitations}
While the work in its current state is helpful for better understanding conversations, it is not without its limitations. Foremost, in the present work we only considered conversations on Reddit. Another limitation is that the entity linking method we chose is geared towards high-precision at the cost of low-recall. This means that we can be confident that the entities extracted in the conversations are mostly correct, but we have missed some portion of entities. The recall limitation does inhibit the total number of entities we were able to collect; a better system would provide for better insights in our downstream analysis. This issue can also be highlighted with the long tail distribution of entities and the challenges this poses to current methods \cite{ilievski2018systematic}. An entity linking model that focuses on recall may still result in useful graphs as prior works have found that many of the entities are considered ``close enough'' even when they are not a perfect match to ground truth data \cite{ding2021posthoc}. Using a different entity linking model could lead to different patterns extracted from our method. For a model that optimizes for higher recall it could create a much larger entity graph, though it would likely contain a fair amount of noise due to the precision-recall trade off.

Another limitation inherent to the present work is the consideration of conversations as threaded trees. This is an imperfect representation of natural, in-person conversation, and still different from unthreaded conversations like those found on Twitter and Facebook, which may require a vastly different entity graph construction method. Finally, the interactive visualization tool is limited in its ability to process enormous amounts of conversation data because of its reliance on JavaScript libraries and interactive browser rendering.

\subsection{Future Work}
These limitations leave open avenues for further exploration in future work. Our immediate goals are to use the entity graphs to better understand how narratives are crafted and shaped across communities. Improvements in the entity linking process and addition of concept vertices, pronoun anaphora resolution, threaded information extraction and other advances in SocialNLP will serve to improve the technology substantially. We also plan to ingest other threaded conversational domains such as Hackernews, 4chan, and even anonymized email data. Extensions of this work could also include capturing more information between entity transitions such as the sentiment  overlayed on a given entity or group of entities. This extra information could allow us to create entity graphs that not only show the transition but also how various groups speak and feel about those specific entities.



\section{Acknowledgements}
The authors would like to thank Yifan Ding and Justus Hibshman for their feedback on the paper.
This work is supported in part by the Defense Advanced Research Projects Agency (DARPA) and Army Research Office (ARO) under Contract No. W911NF-21-C-0002.

\bibliographystyle{acm}
\bibliography{refs}

\begin{thebibliography}{10}

\bibitem{baumgartner2020pushshift}
{\sc Baumgartner, J., Zannettou, S., Keegan, B., Squire, M., and Blackburn, J.}
\newblock The pushshift reddit dataset.
\newblock Proceedings of the international AAAI conference on web and social
  media, 2020.

\bibitem{bostock2011d3}
{\sc Bostock, M., Ogievetsky, V., and Heer, J.}
\newblock D$^3$ data-driven documents.
\newblock {\em IEEE transactions on visualization and computer graphics 17}, 12
  (2011), 2301--2309.

\bibitem{botzer2021reddit}
{\sc Botzer, N., Ding, Y., and Weninger, T.}
\newblock Reddit entity linking dataset.
\newblock {\em Information Processing \& Management 58}, 3 (2021), 102479.

\bibitem{brady2020mad}
{\sc Brady, W.~J., Crockett, M.~J., and Van~Bavel, J.~J.}
\newblock The mad model of moral contagion: The role of motivation, attention,
  and design in the spread of moralized content online.
\newblock {\em Perspectives on Psychological Science 15}, 4 (2020), 978--1010.

\bibitem{centola2010spread}
{\sc Centola, D.}
\newblock The spread of behavior in an online social network experiment.
\newblock {\em science 329}, 5996 (2010), 1194--1197.

\bibitem{chafe1994discourse}
{\sc Chafe, W.}
\newblock {\em Discourse, consciousness, and time: The flow and displacement of
  conscious experience in speaking and writing}.
\newblock University of Chicago Press, 1994.

\bibitem{chafe2017language}
{\sc Chafe, W.}
\newblock Language and the flow of thought.
\newblock {\em The new psychology of language\/} (2017), 93--111.

\bibitem{chandrasekharan2017you}
{\sc Chandrasekharan, E., Pavalanathan, U., Srinivasan, A., Glynn, A.,
  Eisenstein, J., and Gilbert, E.}
\newblock You can't stay here: The efficacy of reddit's 2015 ban examined
  through hate speech.
\newblock {\em Proceedings of the ACM on Human-Computer Interaction 1}, CSCW
  (2017), 1--22.

\bibitem{cheng2013relational}
{\sc Cheng, X., and Roth, D.}
\newblock Relational inference for wikification.
\newblock Empirical Methods in Natural Language Processing, 2013.

\bibitem{cinelli2021echo}
{\sc Cinelli, M., De~Francisci~Morales, G., Galeazzi, A., Quattrociocchi, W.,
  and Starnini, M.}
\newblock The echo chamber effect on social media.
\newblock {\em Proceedings of the National Academy of Sciences 118}, 9 (2021),
  e2023301118.

\bibitem{collins1975spreading}
{\sc Collins, A.~M., and Loftus, E.~F.}
\newblock A spreading-activation theory of semantic processing.
\newblock {\em Psychological review 82}, 6 (1975), 407.

\bibitem{de2017populism}
{\sc De~Cleen, B.}
\newblock Populism and nationalism.
\newblock {\em The Oxford handbook of populism 1\/} (2017), 342--262.

\bibitem{derczynski2015analysis}
{\sc Derczynski, L., Maynard, D., Rizzo, G., Van~Erp, M., Gorrell, G., Troncy,
  R., Petrak, J., and Bontcheva, K.}
\newblock Analysis of named entity recognition and linking for tweets.
\newblock {\em Information Processing \& Management 51}, 2 (2015), 32--49.

\bibitem{ding2021posthoc}
{\sc Ding, Y., Botzer, N., and Weninger, T.}
\newblock Posthoc verification and the fallibility of the ground truth.
\newblock {\em arXiv preprint arXiv:2106.07353\/} (2021).

\bibitem{farrell2019exploring}
{\sc Farrell, T., Fernandez, M., Novotny, J., and Alani, H.}
\newblock Exploring misogyny across the manosphere in reddit.
\newblock Proceedings of the 10th ACM Conference on Web Science, 2019.

\bibitem{fruchterman1991graph}
{\sc Fruchterman, T.~M., and Reingold, E.~M.}
\newblock Graph drawing by force-directed placement.
\newblock {\em Software: Practice and experience 21}, 11 (1991), 1129--1164.

\bibitem{garimella2018political}
{\sc Garimella, K., De~Francisci~Morales, G., Gionis, A., and Mathioudakis, M.}
\newblock Political discourse on social media: Echo chambers, gatekeepers, and
  the price of bipartisanship, 2018.

\bibitem{glenski2019multilingual}
{\sc Glenski, M., Ayton, E., Mendoza, J., and Volkova, S.}
\newblock Multilingual multimodal digital deception detection and
  disinformation spread across social platforms.
\newblock {\em arXiv preprint arXiv:1909.05838\/} (2019).

\bibitem{guntuku2021twitter}
{\sc Guntuku, S.~C., Buttenheim, A.~M., Sherman, G., and Merchant, R.~M.}
\newblock Twitter discourse reveals geographical and temporal variation in
  concerns about covid-19 vaccines in the united states.
\newblock {\em Vaccine 39}, 30 (2021), 4034--4038.

\bibitem{huddy2007american}
{\sc Huddy, L., and Khatib, N.}
\newblock American patriotism, national identity, and political involvement.
\newblock {\em American journal of political science 51}, 1 (2007), 63--77.

\bibitem{ilievski2018systematic}
{\sc Ilievski, F., Vossen, P., and Schlobach, S.}
\newblock Systematic study of long tail phenomena in entity linking.
\newblock Proceedings of the 27th International Conference on Computational
  Linguistics, 2018.

\bibitem{iyengar2019origins}
{\sc Iyengar, S., Lelkes, Y., Levendusky, M., Malhotra, N., and Westwood,
  S.~J.}
\newblock The origins and consequences of affective polarization in the united
  states.
\newblock {\em Annual Review of Political Science 22\/} (2019), 129--146.

\bibitem{jung2020attnio}
{\sc Jung, J., Son, B., and Lyu, S.}
\newblock Attnio: Knowledge graph exploration with in-and-out attention flow
  for knowledge-grounded dialogue.
\newblock Proceedings of the 2020 Conference on Empirical Methods in Natural
  Language Processing (EMNLP), 2020.

\bibitem{kalantari2021characterizing}
{\sc Kalantari, N., Liao, D., and Motti, V.~G.}
\newblock Characterizing the online discourse in twitter: Users’ reaction to
  misinformation around covid-19 in twitter, 2021.

\bibitem{keith2021narrative}
{\sc Keith~Norambuena, B.~F., and Mitra, T.}
\newblock Narrative maps: An algorithmic approach to represent and extract
  information narratives.
\newblock {\em Proceedings of the ACM on Human-Computer Interaction 4}, CSCW3
  (2021), 1--33.

\bibitem{kempe2003maximizing}
{\sc Kempe, D., Kleinberg, J., and Tardos, {\'E}.}
\newblock Maximizing the spread of influence through a social network.
\newblock Proceedings of the ninth ACM SIGKDD international conference on
  Knowledge discovery and data mining, 2003.

\bibitem{kolitsas2018end}
{\sc Kolitsas, N., Ganea, O.-E., and Hofmann, T.}
\newblock End-to-end neural entity linking.
\newblock {\em arXiv preprint arXiv:1808.07699\/} (2018).

\bibitem{kusner2015word}
{\sc Kusner, M., Sun, Y., Kolkin, N., and Weinberger, K.}
\newblock From word embeddings to document distances.
\newblock International conference on machine learning, 2015.

\bibitem{mateas2003narrative}
{\sc Mateas, M., and Sengers, P.}
\newblock {\em Narrative intelligence}.
\newblock J. Benjamins Pub., 2003.

\bibitem{medvedev2017anatomy}
{\sc Medvedev, A.~N., Lambiotte, R., and Delvenne, J.-C.}
\newblock The anatomy of reddit: An overview of academic research.
\newblock Dynamics on and of Complex Networks, 2017.

\bibitem{moon2019opendialkg}
{\sc Moon, S., Shah, P., Kumar, A., and Subba, R.}
\newblock Opendialkg: Explainable conversational reasoning with attention-based
  walks over knowledge graphs.
\newblock Proceedings of the 57th Annual Meeting of the Association for
  Computational Linguistics, 2019.

\bibitem{page2015narrative}
{\sc Page, R.}
\newblock The narrative dimensions of social media storytelling.
\newblock {\em The handbook of narrative analysis\/} (2015), 329--347.

\bibitem{paul2016russian}
{\sc Paul, C., and Matthews, M.}
\newblock The russian “firehose of falsehood” propaganda model.
\newblock {\em Rand Corporation 2}, 7 (2016), 1--10.

\bibitem{ram2018conversational}
{\sc Ram, A., Prasad, R., Khatri, C., Venkatesh, A., Gabriel, R., Liu, Q.,
  Nunn, J., Hedayatnia, B., Cheng, M., Nagar, A., et~al.}
\newblock Conversational ai: The science behind the alexa prize.
\newblock {\em arXiv preprint arXiv:1801.03604\/} (2018).

\bibitem{ran2018attention}
{\sc Ran, C., Shen, W., and Wang, J.}
\newblock An attention factor graph model for tweet entity linking.
\newblock Proceedings of the 2018 World Wide Web Conference, 2018.

\bibitem{rao2021political}
{\sc Rao, A., Morstatter, F., Hu, M., Chen, E., Burghardt, K., Ferrara, E.,
  Lerman, K., et~al.}
\newblock Political partisanship and antiscience attitudes in online
  discussions about covid-19: Twitter content analysis.
\newblock {\em Journal of medical Internet research 23}, 6 (2021), e26692.

\bibitem{schia2020hacking}
{\sc Schia, N.~N., and Gjesvik, L.}
\newblock Hacking democracy: managing influence campaigns and disinformation in
  the digital age.
\newblock {\em Journal of Cyber Policy 5}, 3 (2020), 413--428.

\bibitem{sevgili2020neural}
{\sc Sevgili, O., Shelmanov, A., Arkhipov, M., Panchenko, A., and Biemann, C.}
\newblock Neural entity linking: A survey of models based on deep learning.
\newblock {\em arXiv preprint arXiv:2006.00575\/} (2020).

\bibitem{shahaf2010connecting}
{\sc Shahaf, D., and Guestrin, C.}
\newblock Connecting the dots between news articles.
\newblock Proceedings of the 16th ACM SIGKDD international conference on
  Knowledge discovery and data mining, 2010.

\bibitem{shahaf2013information}
{\sc Shahaf, D., Yang, J., Suen, C., Jacobs, J., Wang, H., and Leskovec, J.}
\newblock Information cartography: creating zoomable, large-scale maps of
  information.
\newblock Proceedings of the 19th ACM SIGKDD international conference on
  Knowledge discovery and data mining, 2013.

\bibitem{shen2014entity}
{\sc Shen, W., Wang, J., and Han, J.}
\newblock Entity linking with a knowledge base: Issues, techniques, and
  solutions.
\newblock {\em IEEE Transactions on Knowledge and Data Engineering 27}, 2
  (2014), 443--460.

\bibitem{tadesse2019detection}
{\sc Tadesse, M.~M., Lin, H., Xu, B., and Yang, L.}
\newblock Detection of depression-related posts in reddit social media forum.
\newblock {\em IEEE Access 7\/} (2019), 44883--44893.

\bibitem{van2020rel}
{\sc van Hulst, J.~M., Hasibi, F., Dercksen, K., Balog, K., and de~Vries,
  A.~P.}
\newblock Rel: An entity linker standing on the shoulders of giants, 2020.

\bibitem{weedon2017information}
{\sc Weedon, J., Nuland, W., and Stamos, A.}
\newblock Information operations and facebook.
\newblock {\em Retrieved from Facebook: https://fbnewsroomus. files. wordpress.
  com/2017/04/facebook-and-information-operations-v1. pdf\/} (2017).

\bibitem{yu2020survey}
{\sc Yu, W., Zhu, C., Li, Z., Hu, Z., Wang, Q., Ji, H., and Jiang, M.}
\newblock A survey of knowledge-enhanced text generation.
\newblock {\em arXiv preprint arXiv:2010.04389\/} (2020).

\bibitem{zhang2019grounded}
{\sc Zhang, H., Liu, Z., Xiong, C., and Liu, Z.}
\newblock Grounded conversation generation as guided traverses in commonsense
  knowledge graphs.
\newblock {\em arXiv preprint arXiv:1911.02707\/} (2019).

\bibitem{zien1999multilevel}
{\sc Zien, J.~Y., Schlag, M.~D., and Chan, P.~K.}
\newblock Multilevel spectral hypergraph partitioning with arbitrary vertex
  sizes.
\newblock {\em IEEE Transactions on computer-aided design of integrated
  circuits and systems 18}, 9 (1999), 1389--1399.

\bibitem{zomick2019linguistic}
{\sc Zomick, J., Levitan, S.~I., and Serper, M.}
\newblock Linguistic analysis of schizophrenia in reddit posts.
\newblock Proceedings of the Sixth Workshop on Computational Linguistics and
  Clinical Psychology, 2019.

\end{thebibliography}

\end{document}